\documentclass[lettersize,journal]{IEEEtran}
\usepackage{amsmath,amsfonts,amsthm}
\usepackage{algorithmic}
\usepackage{algorithm}
\usepackage{array}
\usepackage[caption=false,font=normalsize,labelfont=sf,textfont=sf]{subfig}
\usepackage{textcomp}
\usepackage{stfloats}
\usepackage{url}
\usepackage{booktabs}
\usepackage{verbatim}
\usepackage{graphicx}
\usepackage{cite}
\usepackage[T1]{fontenc}
\graphicspath{{images/}} 
\usepackage{xcolor}
\newcommand{\blue}[1]{\textcolor{black}{#1}}

\theoremstyle{remark}
\newtheorem{remark}{Remark}

\begin{document}

\title{BMG-Q: Localized Bipartite Match Graph Attention Q-Learning for Ride-Pooling Order Dispatch}

\author{Yulong Hu, Siyuan Feng, and Sen Li
\thanks{This work was supported by the Hong Kong Research Grants Council under project 16202922, National Natural Science Foundation of China under project 72201225, and General Research Fund from the Research Grants Council of the Hong Kong Special Administrative Region  under Project PolyU15207424. {(\em co-Corresponding author: Sen Li and Siyuan Feng)}} %
\thanks{Y. Hu and S. Li are with the Department of Civil and Environmental Engineering, The Hong Kong University of Science and Technology, (email: yulong.hu@connect.ust.hk, cesli@ust.hk), S. Li is also affiliated with Intelligent Transportation Thrust, Systems Hub, The Hong Kong University of Science and Technology (Guangzhou), S. Feng is with the Department of Aeronautical and Aviation Engineering, The Hong Kong Polytechnic University (email:siyuan.feng@polyu.edu.hk).}
\thanks{This work has been published in IEEE Transactions on Intelligent Transportation Systems, DOI: 10.1109/TITS.2025.3595653.}

}

\markboth{IEEE Transactions on Intelligent Transportation Systems}%
{Yulong \MakeLowercase{\textit{et al.}}: BMG-Q: Towards Effective, Scalable, and Robust Ride-Pooling Dispatch with Localized Bipartite Match Graph Attention Q-Learning}


\maketitle

\begin{abstract}
This paper introduces Localized Bipartite Match Graph Attention Q-Learning (BMG-Q), a novel Multi-Agent Reinforcement Learning (MARL) algorithm framework tailored for ride-pooling order dispatch. BMG-Q advances ride-pooling decision-making process with the localized bipartite match graph underlying the Markov Decision Process, enabling the development of novel Graph Attention Double Deep Q Network (GATDDQN) as the MARL backbone to capture the dynamic interactions among ride-pooling vehicles in fleet. Our approach enriches the state information for each agent with GATDDQN by leveraging a localized bipartite interdependence graph and enables a centralized global coordinator to optimize order matching and agent behavior using Integer Linear Programming (ILP). Enhanced by gradient clipping and localized graph sampling, our GATDDQN improves scalability and robustness. Furthermore, the inclusion of a posterior score function in the ILP captures the online exploration-exploitation trade-off and reduces the potential overestimation bias of agents, thereby elevating the quality of the derived solutions. Through extensive experiments and validation, BMG-Q has demonstrated superior performance in both training and operations for  thousands of vehicle agents, outperforming benchmark reinforcement learning frameworks by around 10\% in accumulative rewards and showing a significant reduction in overestimation bias by over 50\%. Additionally, it maintains robustness amidst task variations and fleet size changes, establishing BMG-Q as an effective, scalable, and robust framework for advancing ride-pooling order dispatch operations.
\end{abstract}


\section{Introduction}
\IEEEPARstart{T}{he} widespread adoption of mobile communication and Global Positioning System technology has allowed Transportation Network Companies (TNCs) such as Uber, Lyft, and Didi to provide on-demand mobility services on a global scale~\cite{jiang2018ridesharing,tong2017simpler}. Ever since~\cite{alonso2017demand}, the advantages of flexible and collaborative ride-sharing operations have become increasingly recognized within the transportation research community. In line with this trend, there has been an expanding body of research on operational policies for ride-sharing, including the coordination of ride-hailing with public transportation~\cite{feng2022coordinating}, ride-sharing with passenger transfers~\cite{wang2023optimization}, and the coordination of autonomous vehicles with conventional vehicles~\cite{xie2023two}. These advancements are propelled by breakthroughs in deep learning and Multi-Agent Reinforcement Learning (MARL) frameworks~\cite{berner2019dota,ouyang2022training}.

Nevertheless, several hurdles must be overcome to unlock the full potential of MARL for developing effective, scalable, and robust real-time operational strategies for ride-pooling order dispatch. A principal challenge in this context is the complex interdependence in decision-making among vehicles, which leads to an exponential increase in both state and action spaces within large fleets \cite{sutton1998introduction}.  One approach to address this is by traditional independent learning approaches, such as Independent Q-Learning (IQL) and Independent Proximal Policy Optimization (IPPO)~\cite{de2020independent,tampuu2017multiagent}, which ignore the interdependence. In the context of ride-pooling, it is common in the existing literature to  combine single-agent independent Reinforcement Learning (RL) with bipartite matching. For instance, \cite{al2019deeppool} proposed to adopt a Deep Q-Network (DQN) for relocating ride-pooling agents and bipartite matching for order dispatch. To improve the transferability and scalability of the framework, \cite{sadeghi2022reinforcement} introduced additional techniques such as limited-memory upper confidence bound and reward smoothing. Moreover, \cite{feng2022coordinating} coordinated ride-hailing with public transit by encoding the decisions of subway stations into the states of tabular temporal difference learning. Similarly, \cite{wang2023optimization} facilitated ride-pooling with passenger transfer. Yet, the practice of merging independent reinforcement learning with bipartite matching, while improving scalability, often overlooks the agents' complex interdependence during the RL exploitation and training phases. This can lead to significant overestimation of rewards, a critical concern in highly competitive environments such as ride-pooling, where the pronounced interdependence among agents intensifies the issue.

To accommodate the intricate interdependence among agents, several MARL frameworks have been introduced, including state-of-the-art Centralized Training with Decentralized Execution (CTDE) algorithms such as Multi-Agent Deep Deterministic Policy Gradient (MADDPG) \cite{lowe2017multi}, Q-mix \cite{rashid2020monotonic}, Q-tran \cite{son2019qtran}, and Multi-Agent Proximal Policy Optimization (MAPPO) \cite{yu2022surprising}. However, these methods are typically applied to much smaller-scale problems. In contrast, large-scale ride-pooling order dispatch involves thousands of agents, rendering these approaches infeasible. To enhance algorithmic scalability while capturing agent interactions, researchers have explored novel concepts such as Mean-Field MARL \cite{yang2018mean}, where agents interact with an average representation of all other agents. However, in the ride-pooling context, this average state may not accurately represent agent interdependence as each agent has unique passengers with different itineraries, and summing this up may lead to misleading information for the decision maker.  One approach to overcome this limitation is by considering the Attention-based MARL~\cite{iqbal2019actor,jiang2018learning}, which instead of relying on average state information, allows distinct neighboring agent to has distinct weights that can be flexibly adjusted over time. Nevertheless, the application of Attention-based MARL is limited in small-scale scenarios and single-passenger ride-hailing \cite{jiang2018learning,kullman2022dynamic}, where the possibility of multiple riders sharing the same ride is not explicitly considered.  

To fill the above-mentioned research gaps, we introduce a novel MARL framework specifically crafted for ride-pooling order dispatch—the Localized \textbf{B}ipartite \textbf{M}atch \textbf{G}raph Attention \textbf{Q}-Learning (BMG-Q). This framework is adept at capturing the intricate interdependencies that typically arise within a localized graph, defined by the bipartite matching radius. By implementing the Graph Attention Double Deep Q-Network (GATDDQN), we provide ride-pooling agents with enriched state representations that factor in the influence of other agents' actions during decision-making. Techniques such as gradient clipping and graph sampling have been employed to bolster the robustness and scalability of the GATDDQN, ensuring that agents retain learned information over time rather than overfitting to recent transitions. In addition, we have seamlessly integrated the GATDDQN with a bipartite matching mechanism through a posterior score function and Integer Linear Programming (ILP). This integration enhances the central matcher's efficiency and refines the balance between exploration and exploitation. Our comprehensive numerical studies show that BMG-Q can effectively model the complex interactions among agents, reduce overestimation bias, and improve overall performance. The study also verifies that BMG-Q retains its robustness in the face of task variability and training hyper-parameter changes,  thus establishing it as an effective, scalable, and robust approach for advancing ride-pooling operations. The major contributions of this paper are summarized below:
\begin{itemize}
    \item We propose a novel BMG-Q framework to address multi-agent interactions in MARL within the context of large-scale ride-pooling order dispatch. The propsoed framework leverages the novel localized bipartite match interdependent Markov Decision Proces (MDP) formulation with the Graph Attention Double Deep Q Network (GATDDQN) as backbones. It captures the interdependence among agents and thus leads to more effective assignment decisions compared to existing works.
    \item Our work stands at the forefront of developing graph-based MARL techniques for large-scale ride-pooling order dispatch systems. While contemporary studies in the realm of MARL have started to explore the incorporation of GNN with RL \cite{jiang2018graph,liu2020multi, bohmer2020deep,wei2022vgn,munikoti2023challenges}, they often encounter limitations due to scalability, stability, and robustness. By employing strategic measures such as gradient clipping and random graph sampling, our BMG-Q framework showcases a consistently robust training and validation performance in systems comprising thousands of agents and the face of task variations and parameter changes.
    \item We validate the BMG-Q framework through a case study in New York City, utilizing a real-world taxi trip dataset \cite{NYCTaxiData2018, ProjectOSRM}. We demonstrate that our proposed framework not only significantly reduces overestimation issues but also outperforms benchmark frameworks. This is evidenced by an approximate 10\% increase in total accumulated rewards and a more than 50\% reduction in overestimation, underscoring the enhanced performance of our BMG-Q ride-pooling dispatch operations.
    
\end{itemize}

\section{Related Works}
\textbf{Ride-Pooling Order Dispatch.} The operational dynamics of ride-pooling have garnered considerable attention due to their promising yet unpredictable real-time demand, as evidenced by various studies\cite{alonso2017demand}. The nature of this uncertainty, coupled with the full potential of ride-pooling systems, introduces complexity into the process of coordinating vehicles with multiple passengers. Effective coordination requires not only addressing the needs of current passengers but also anticipating the needs of future riders, which includes managing new ride requests and those already being served. Initial investigations in this field have considered short-sighted, or myopic, policies that make vehicular assignments based on presently available information \cite{alonso2017demand, simonetto2019real}. Specifically, \cite{alonso2017demand} notably advances this by introducing the shareability graph, which identifies possible sharing opportunities between new requests and vehicles on standby. They put forward a batch-matching strategy and crafted a sequential method that divides the decision-making process into vehicle routing and passenger assignment tasks. For more efficient real-time operations, \cite{simonetto2019real} reduces the complexity of the matching problem by limiting the process to pairing a single passenger with a vehicle at each time step. More recent advancements in the field have shifted towards a better incorporation of the uncertainties related to future demand into the decision-making processes via methods such as model predictive control\cite{ali2023rebalancing}, approximate dynamic programming\cite{you2024approximate}, and stochastic integer programming\cite{luo2023efficient}. Note that these works are model-based, requiring explicit characterization of  system dynamics and/or future uncertainties. 

\textbf{MARL Framework for Ride-Pooling Dispatch.} Given the super-human capabilities of RL and MARL showcased in a range of notable achievements~\cite{mnih2015human,berner2019dota,ouyang2022training}, the prospect of crafting practical MARL systems for the real-time optimization of ride-sharing dispatch grows increasingly compelling. While some researchers have endeavored to deploy multi-agent reinforcement learning approaches such as Mean-Field MARL \cite{zhu2021mean}, Q-mix \cite{de2020efficient}, and Attention-based MARL \cite{kullman2022dynamic,enders2023hybrid} in ride-sourcing scenarios, these methods continue to grapple with challenges like stability and scalability when it comes to training in large-scale and complex settings. To address the scalability issue in large-scale ride-sharing systems, it is common in the ride-sharing research community to combine single agent RL (or equivalently, Independent RL) with bipartite match. Specifically, \cite{al2019deeppool} propose to adopt DQN for ride-pooling agents' relocation and bipartite match for order dispatch. To improve the transferability and scalability of full deployment of the framework in ride hailing, \cite{sadeghi2022reinforcement} proposes additional techniques such as limited-memory upper confidence bound and reward smoothing. Moreover, \cite{feng2022coordinating} coordinates ride-sourcing with public transit through encoding the decision of subway stations into states of tabular temporal difference learning. \cite{xie2023two} coordinates autonomous vehicles with conventional vehicles through two-sided deep reinforcement learning. \cite{wang2023optimization} enable ride-pooling with passenger transfer. However, the aforementioned works have yet to adequately address the intricate interdependencies among vehicles in the ride-pooling context while also ensuring scalability for MARL.

\textbf{Graph-based MARL.} As the computational efficiency and representational power of GNN models such as Graph Convolutional Network (GCN)~\cite{kipf2016semi}, GraphSAGE~\cite{hamilton2017inductive}, and Graph Attention Network (GAT)~\cite{velivckovic2017graph} gain increasing recognition in complex and adaptive representation learning, researchers have begun to investigate the integration of these potent GNN models with MARL. This nascent area of research seeks to tackle a variety of challenges within MARL, such as the complex task of encoding environmental dynamics from the perspective of individual agents, as well as the decomposition of value functions and the nuanced distribution of credit across the collective team~\cite{munikoti2023challenges}. Specifically for coordination games, on the one hand, \cite{jiang2018graph,liu2020multi} propose to adopt graph convolution RL and two-stage attention mechanism to learn abstract interplay representation between agents within graph topology. \cite{wei2022vgn} propose the idea of coordination graph and utilize GATs to factorize the join team value function or team policy to enable coordination behavior among agents. Despite these advancements, these advances have not yet been applied to ride-pooling, a highly complex and large-scale system, where achieving scalability, stability, and robustness concurrently remains a significant challenge.

\section{Problem Formulation \& Benchmark Methods}
In this section, we formulate the ride-pooling order dispatch problem and review the strategies commonly used in previous literature. The ride-pooling vehicles are conventionally considered as independent and homogeneous agents under the bipartite matching process. A benchmark method will be established, against which we can compare our proposed algorithm in subsequent discussions.

In particular, we will begin by presenting the MDP formulation for ride-pooling order dispatch, detailing each agent's state, action, reward, discount factor, and transition function under various scenarios in Subsection A. We then explore how the assumptions of independence and homogeneity, prevalent in the ride-pooling community's approach, serve to decentralize the original MDP. Building upon the analysis, we illustrate the integration of Independent RL with ILP  and then outline how RL techniques, such as Double Deep Q-Network (DDQN), could be applied to learn and represent the system's dynamics to finally form a benchmark framework, termed ILPDDQN, for ride-pooling order dispatch.

\subsection{MDP Formulation for Ride-Pooling Order Dispatch}
The ride-pooling order dispatch problem is normally formulated as a multi-agent MDP, with each ride-pooling vehicle representing an agent (we refer to it as agent or vehicle agent hereafter).  Each vehicle agent's definition of state, action, reward, and transition function could be detailed as follows:

{
\setlength{\parindent}{0pt}
1) \textit{\textbf{State}}: For each vehicle $n$ at time $t$, its state is $s_{n,t} = (l_{n,t}, v_{n,t}, p_{n,t},$ $ o_{n,t}, d_{n,t}, t)$, where $l_{n,t}$ encodes the current location; $v_{n,t}$ is the number of vacant seats; $p_{n,t}$ encodes the information of passengers on board, including their estimated remaining time on board, drop-off locations, and current additional travel time; $o_{n,t}$ and $d_{n,t}$ represent a set of origin and destination pairs of the observed incoming orders within the matching distance of agent $n$, respectively; and $t$ is the current time.

2) \textit{\textbf{Action}}: For available vehicle $n$ (i.e., the vehicle is not full or in the process of picking up a new passenger) at time $t$, after seeing the incoming new orders, platform assigns action $a_{n,t}$ to decide whether to pick up one of the observed passengers: if not, we have $a_{n,t} = 0$ and then vehicle $n$ will remain idle or continue with the remainder of its trip as determined by the on-board passenger's itinerary; otherwise if the vehicle is assigned by the platform to pick up the $z^{th}$ request (among all observed incoming orders of vehicle $n$), then we have $a_{n,t} = z$. 

3) \textit{\textbf{Reward Function}}: If vehicle $n$ is not available or does not accept any of the observed new order at time $t$, then the reward function at time $t$ is as:
\begin{eqnarray}
r_{n,t}(s_{n,t},a_{n,t}) = -c_0,
\end{eqnarray}
where $c_0$ is the cost of the vehicle, including both operational cost and amortized capital cost. If vehicle $n$ accepts any of the observed new order, then the reward function can be written as:
\begin{equation}
\begin{aligned}
r_{n,t}(s_{n,t},a_{n,t}) &= \beta_{0} + \beta_{1} \cdot Dis \\
                                &-\beta_{2}\cdot Pickup \\
                                &- \beta_{3} \cdot min(Add,thre) \\
                                &- \beta_{4} \cdot max(Add-thre, 0)-c_0
\end{aligned}
\end{equation}
where the first term is the starting revenue of a vehicle picking up a new passenger; the second term is the revenue based on the distance between between new order’s origin and destinations (denoted as $Dis$);  the third term is the cost of the new passenger waiting to be picked up (with waiting time denoted by $Pickup$); the intuition of the fourth and fifth terms is to give a small penalty if the total additional travel time due to ride pooling compared with a direct non-sharing ride-hailing trip (denoted as $add$) is below the threshold time (denoted as $thre$) but give a heavy penalty if the additional time is above the threshold.
For the platform as a whole, the total reward $R_t$ at time $t$ is the summation of rewards of all $N$ agents at time $t$:
\begin{eqnarray}
{R_t}(S_{t},A_{t}) = \sum_{n = 1}^{N} r_{n,t}(s_{n,t},a_{n,t})
\end{eqnarray}
where state $S_t$ and action $A_t$ at time $t$ are respectively the collections of state and action of all $N$ agents at time $t$:
\begin{eqnarray}
{S_t} = [{s_{1,t}},{s_{2,t}}, \ldots ,{s_{N,t}}]
\end{eqnarray}
\begin{eqnarray}
{A_t} = [{a_{1,t}},{a_{2,t}}, \ldots ,{a_{N,t}}]
\end{eqnarray}
4) \textit{\textbf{State Transition Function}}: The transition function can be represented in the form of $P(S_{t+1}|S_{t},A_{t})$. The explicit form of $P(\cdot|\cdot, \cdot)$ and reward function $R(\cdot, \cdot)$ is unknown and will be learned later via RL/MARL methods.

}

\begin{figure*} [ht]
  \centering
  \includegraphics[width=0.7\linewidth]{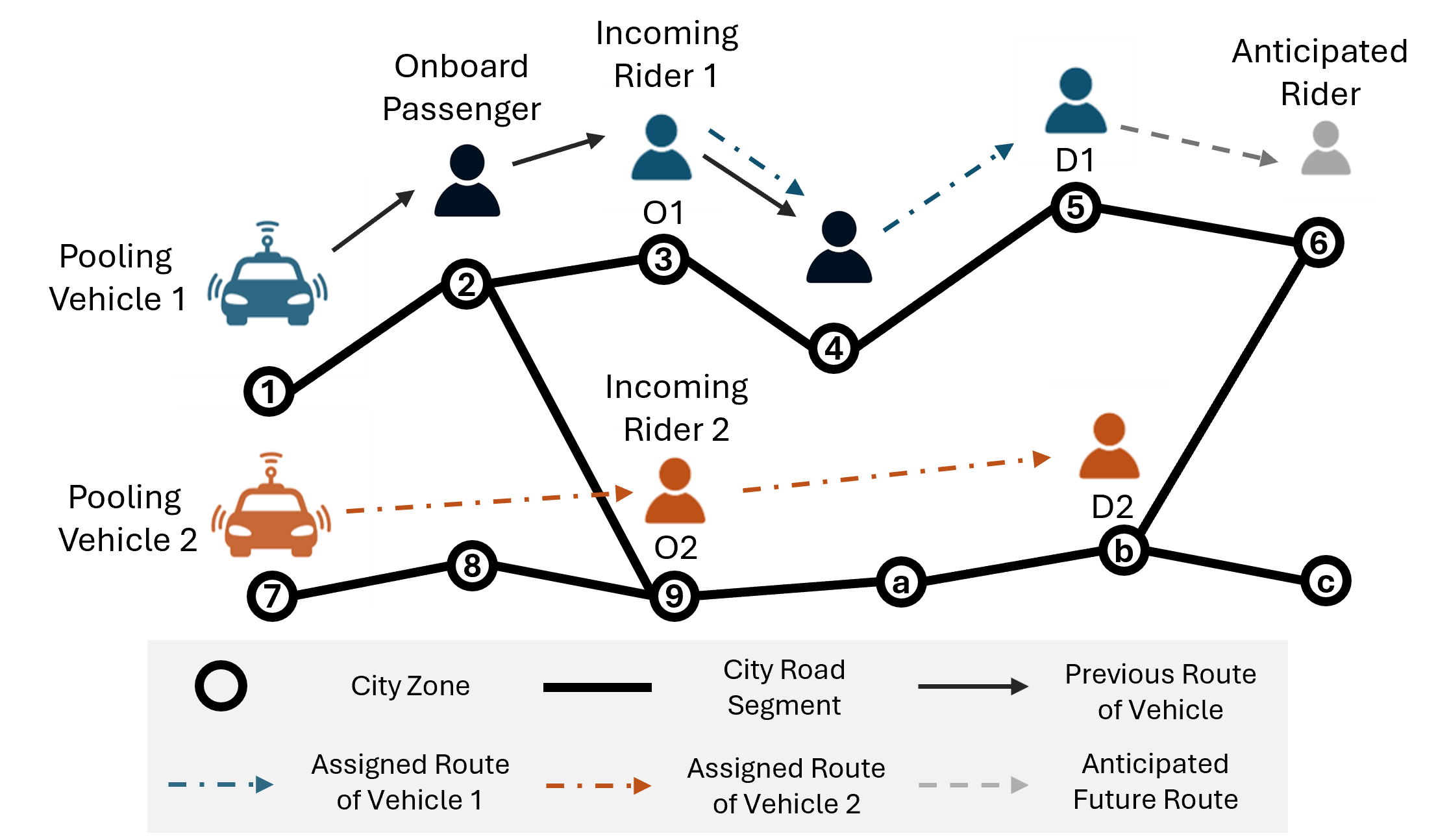}
  \caption{An illustrative example of decision-making process for two agents. At time $t$, two new requests are observed, including (a) Rider 1 from zone 3 to zone 5;  and (b) Rider 2 from zone 9 to zone b. The platform then collaborative dispatches two vehicles: Vehicle 1 is assigned with action $a_{1,t}=1$ to integrate Rider 1 into its current route; while Vehicle 2 is idle and dispatched with action $a_{2,t} = 2$ to pickup Rider 2.}
  \vspace{-1em}
  \label{fig:Visualization of Ride-pooling Agent's Decision Process}
\end{figure*}

To further delineate the decision-making process of agents in ride-pooling scenarios more clearly, we refer to the illustrative example presented in Figure~\ref{fig:Visualization of Ride-pooling Agent's Decision Process}, which features two collaborative agents. At time $t$, the figure displays two pooling vehicles, Vehicle 1 and Vehicle 2, awaiting dispatch decisions. Vehicle 1 is already committed to picking up a passenger from zone 2 and dropping him/her off at zone 4, while Vehicle 2 is idle at the moment. Two new requests are observed, including: (a) Rider 1 from zone 3 to zone 5; and (b) Rider 2 from zone 9 to zone b. The platform then collaborative dispatches two vehicles: Vehicle 1 is assigned action $a_{1,t}=1$, to pick up Rider 1. This action is integrated seamlessly with its current route, optimizing the journey for the existing passenger and enhancing operational efficiency. Concurrently, Vehicle 2 is designated action $a_{2,t} = 2$, to pick up Rider 2, effectively utilizing its idle status. With assistance of RL methods, the collaborative decisions should enable Vehicle 1 to address the immediate needs of its onboard passenger while also strategically planning for an anticipated future pickup in zone~6.

\subsection{Independent and Homogeneous Assumptions in MDP}
Consider a ride-pooling platform with $N$ vehicles. At time $t$, agent $n$ can observe its own state $s_{n,t}$ and chooses action $a_{n,t}$. The Q-value of the overall platform (encompassing all the vehicles), represented as $Q_{tot}(S_t, A_t)$, could be expressed as:
\begin{equation}
\begin{aligned}\label{eq:orignal Qtotal}
 Q_{\text{tot}}(S_t, A_t) = \mathbb{E}_{\Pi} \left[ \sum_{k = 0}^{\infty} \gamma^k R_{t + k + 1} \mid S_t, A_t \right],
\end{aligned}
\end{equation}
where $\Pi(\cdot)$ is the centralized policy that maps the state space to the action space, $\gamma$ is the discounted factor, $R_t$ is the joint reward of all agents at time $t$. 

The objective for the platform is to find the optimal policy that maximizes the joint expected discounted cumulative reward over time, which could be expressed as: 
\begin{eqnarray}
{Q_{\text{tot}}^*}({S_t},{A_t}) = \mathbb{E}_{\Pi^*} \left[ \sum_{k = 0}^{\infty} \gamma^k R_{t + k + 1} \mid S_t, A_t \right]
\end{eqnarray}
where $\Pi(\cdot)^*$ is the centralized optimal policy, a mapping from the state space to the action space. 

However, in ride-pooling, thousands of agents might need to be simultaneously dispatched, which will lead to  a prohibitively high dimensional state and action space for the above centralized MDP. Encountering this challenge, it is common in the previous literature to assume agents are \textit{\textbf{independent}} \cite{al2019deeppool,sadeghi2022reinforcement, feng2022coordinating, wang2023optimization}, i.e., each agent’s transition function and reward function has no interdependence with other agents’ actions (note that our proposed method does not require this assumption), which largely reduces the MDP dimensionality and decentralizes the original transition function from $P(S_{t+1}|S_{t},A_{t})$ to $p(s_{t+1}|s_{t},a_{t})$. The independent assumption modifies the centralized MDP in Equation~(\ref{eq:orignal Qtotal}) into:
\begin{equation}
\begin{aligned} \label{eq:seq independent Qtotal}
Q_{\text{tot}}(S_t, A_t) &= \sum_{n = 1}^{N} {\mathbb{E}_{\pi_n} \left[ \sum_{k = 0}^{\infty} \gamma^k r_{n,t + k + 1} \mid s_{n,t}, a_{n,t} \right] }\\
&  = \sum\limits_{n = 1}^N {{Q_n}({s_{n,t}},{a_{n,t}}} )
\end{aligned}
\end{equation}
where $\pi_n$ is the individual policy held by agent $n$ and $r_{n,t}$ is the reward of agent $n$ at time $t$, and $Q_n$ is defined as the expected accumulative reward of agent $n$ under $s_{n,t}, a_{n,t}$.

Furthermore, by assuming the vehicle agents are \textit{\textbf{homogeneous}} (which is often the case in TNCs) and share the same policy $\pi$, Equation~(\ref{eq:seq independent Qtotal}) could be further simplified into:
\begin{equation}
\begin{aligned}
Q_{\text{tot}}(S_t, A_t) &= \sum_{n = 1}^{N} {\mathbb{E_{\pi}} \left[ \sum_{k = 0}^{\infty} \gamma^k r_{n,t + k + 1} \mid s_{n,t}, a_{n,t} \right] }\\
&  = \sum\limits_{n = 1}^N {{Q}({s_{n,t}},{a_{n,t}}})
\end{aligned}
\end{equation}
Then, the goal of the simplified MDP model is to find the optimal policy $\pi^*$ that maximizes the joint expected discounted cumulative reward over time:
\begin{equation}
Q^*_{\text{tot}}({S_t},{A_t}) =  \sum\limits_{n = 1}^N {\mathbb{E}_{\pi^*} \left[ {\sum\limits_{k = 0}^\infty  {{\gamma ^k}{r_{n,t + k + 1}}|{s_{n,t}},{a_{n,t}}} } \right]} 
\end{equation}

\subsection{Merging Independent Learning with Bipartite Matching}
Since simply adopting independent assumptions will make the system ignore the complexity compounded by the likelihood of agents making concurrent decisions and fall into conflicts of agents accepting the same order, previous research has often employed a hybrid approach that melds independent reinforcement learning's policy function or value function with a bipartite matching process \cite{al2019deeppool, sadeghi2022reinforcement, feng2022coordinating, wang2023optimization}. Here we adopt the representative integration of ILP with Independent value function like $Q(s,a)$ to illustrate this idea. 

At each matching time window, the central platform calculates the estimated cumulative total rewards for every feasible agent-order pair and determines the optimal order assignment to vehicles to maximize the platform's overall profit. In this scenario, the platform's objective when making assignment decisions can be approximated by the sum of all Q-values. The optimal assignment problem can thus be formulated as an ILP problem, as presented in Equation (\ref{eq:ILP}) below:
\begin{equation} \label{eq:ILP} 
\begin{aligned} 
& \underset{x_{i,j}}{\text{maximize}} 
& & Z(X) = \sum_{n=1}^{N} \sum_{z=0}^{Z_{t}} Q(s_{n,t},z) x_{n,z} \\ 
& \text{subject to} 
& & \sum_{n=1}^{N} x_{n,z} \leq 1, \quad \forall z, \\ 
& & & \sum_{z=1}^{Z_{t}} x_{n,z} \leq 1, \quad \forall n, \\ 
& & & x_{n,z} \in \{0,1\}, \quad \forall n,z \\
& & & \sum_{n=1}^{N} x_{n,z}\cdot d_{n,z}\leq R_{match}, \quad \forall z, \\
\end{aligned} 
\end{equation}
where $N$ is the total number of vehicles, $Z_{t}$ is the total number of observed orders by the platform at time $t$, $x_{n,z}$ denotes the matching decision for a specific vehicle-order pair, and $d_{n,z}$ is the distance between vehicle $n$ and order $z$. This formulation is subject to constraints ensuring that at each decision time window, a vehicle (e.g., vehicle $n$) can only be matched with one order within the matching distance $R_{match}$ (e.g., such as order $z$), and similarly, one order can only be matched with one vehicle within this matching radius. 

\begin{remark}
In our study, we adopt the common assumption consistent with many existing literature: each vehicle is assigned only one request per time period. This aligns with many established methodologies, as seen in \cite{al2019deeppool,haliem2021adapool,liu2022deep,enders2023hybrid}. Note that this assumption does not impose a significant loss of optimality compared to assigning bundled orders to the same vehicle simultaneously \cite{alonso2017demand}. Specifically, in our context, dispatch decisions are made very frequently (e.g., every minute) in a dynamic manner. If the system intends to assign multiple requests to the same vehicle, it can first assign one order at the current time step, and even before the first order is picked up, it can assign another order to the same vehicle in a subsequent time period. This approach can actually be more efficient and effective than assigning two orders to the same vehicle simultaneously. This is because deferring bundling decisions to future time points, when new information may become available, allows the platform greater flexibility to dynamically adjust decisions under uncertainties.
\end{remark}

\subsection{ILPDDQN Benchmark Framework}
To learn the dynamics of the environment under the above-formulated framework,  we will first review DDQN~\cite{van2016deep} as the backbone structure to learn reward and transition function from vehicle trajectories with format as ($s_i, a_i, r_i, s_i'$), where $s_i$ is the current state of vehicle $i$, $a_i$ is the action taken by vehicle $i$, $r_i$ is the reward received by vehicle $i$, and $s_i'$ is the next state of vehicle $i$. 
Compared with DQN~\cite{mnih2015human}, DDQN manages to mitigate the overestimation of Q-value by using the training network to select the best action for the generation of TD target in the loss calculation during training update, which could be formulated as follows:
\begin{equation} \label{eq:DDQN Loss Calculation}
\begin{aligned}
L = \mathbb{E}_{\tau \sim \mathcal{D}}\Bigg[&\Bigg{(}r_i + \gamma Q\left(s_i',\mathop{\arg\max}_{a_i'} Q(s_i',a_i';\theta);\theta^- \right) \\
&- Q(s_i,a_i;\theta)\Bigg{)}^2\Bigg] 
\end{aligned}
\end{equation}
where $Q(s,a;\theta)$ is the Q value estimated by the training network whose neural network parameter is $\theta$ and $Q(s,a;\theta^-)$ is the Q value estimated by the target network $\theta^-$, $\tau$ is the trajectory from the sampled mini-batch $\mathcal{D}$.
The parameters of DDQN training network will be updated through gradient descent with the equation as follows, where $\alpha$ is the learning rate. 
\begin{equation} \label{eq:DDQN update}
\theta = \theta  - \alpha {\Delta _\theta } L
\end{equation}
For updating DDQN target network, Polyak Average is popular to be adopted for soft update \cite{fujimoto2018addressing} to map training network parameters to target network parameters after every training step as follows to help to stabilize the training process:
\begin{equation} \label{eq:Polyak Average}
\theta^- = \rho \cdot \theta + (1 - \rho) \cdot \theta^-
\end{equation}
where $\rho$ is the soft update hyper-parameter.

Moreover, to encourage exploration and exploitation trade-off at the early stage of the game, exploration decay and epsilon-greedy policy are adopted in DDQN. The formulation of exploration decay and epsilon-greedy policy is given in Equations (\ref{eq:epsilon_decay}) and (\ref{eq:epsilon_greedy_policy}) respectively, where $\epsilon$ is the current exploration rate, $\beta$ is the decay rate, and $\epsilon_T$ is a small predefined threshold exploration rate. DDQN still also employs experience replay to break the correlation of sequential experiences. This is a critical feature that prevents the update process from becoming cyclical and counterproductive, ensuring a more stable and effective learning progression.

\begin{equation} \label{eq:epsilon_decay}
\epsilon = \max(\epsilon \cdot \beta, \epsilon_T)
\end{equation}
\vspace{-1em}
\begin{equation} \label{eq:epsilon_greedy_policy}
\pi^*(s_t)= 
\begin{cases} 
\mathop{\arg\max}\limits_{a_{n,t}} Q^*(s_{n,t},a_{n,t}), & \text{with prob } 1 - \epsilon \\
\text{a random action}, & \text{with prob } \epsilon
\end{cases}
\end{equation}
\vspace{-1em}

\begin{algorithm} 
\caption{ILPDDQN Framework} \label{al: ILPDDQN Framework}
\begin{algorithmic}[1] 
\STATE Simulator Initialization: Episode Order Requirements, Open Street Routing Mmaching (OSRM) Router Model\cite{ProjectOSRM}, Matching Distance $R_{match}$, Number of Vehicles $N$.
\STATE DDQN Initialization: Memory $M$, Memory Capacity $C$, Training Net Parameter $\theta$, Target Net Parameter $\theta^-$, and Training Hyper-parameters $\alpha$, $\rho$, Exploration Rate $\epsilon$, $\epsilon_T$ with Exponential Decay Rate $\beta$.
\FOR{$e = 1$ to Episodes}
    \STATE Initialize: Episode Order Requirements, and Number of Vehicles $N$.
    \FOR{$t = 0$ to $t_{\text{terminal}}$ by $\Delta t$}
        \STATE Central platform updates order information, each vehicle’s location, and on-board passenger situations.
        \STATE Central platform assigns orders to vehicle agents according to ILP formulation in Equation (\ref{eq:ILP}) with the value estimation of the training network.
        \STATE Vehicles observe their orders and perform the assigned actions in the simulation platform and add every agent’s new experience tuple $(s, a, r, s')$ into the memory $M$.
        \IF{memory size larger than $C$}
            \STATE Sample $N$ experience tuples $(s, a, r, s')$ in $M$ as mini-batch $D$ and use Equation~(\ref{eq:DDQN update}) to update $\theta$.
            \STATE Update target network parameters $\theta^-$ using Equation~(\ref{eq:Polyak Average}).
        \ENDIF
        \STATE Based on the chosen action, central platform calculates the new route and estimated time of pickup and drop off.
    \ENDFOR
\ENDFOR
\end{algorithmic}
\end{algorithm}
Overall,  we could assemble the MARL with bipartite matching to establish a {\em benchmark} algorithmic framework, terms as ILPDDQN, which is detailed in Algorithm \ref{al: ILPDDQN Framework}. In particular, after initialization simulator and DDQN networks in step 1 to 4, at every time window of the episode, central platform (matcher) will firstly update order information in step 6 and then perform bipartite match according to ILP calculations in step 7. Under bipartite match, vehicle agents will observe the matched orders, perform order choice actions, and then finally update DDQN network parameters from step 8 to 11. However, it is worth noting that now the experiences are collected and shared by every agent due to homogeneity and independence assumptions.

\section{Localized Bipartite Match Interdependent MDP and GATDDQN}
While the previously introduced framework that combines independent RL with bipartite matching (i.e., ILPDDQN) significantly enhances scalability, it overlooks the intricate interdependence among agents in the MARL exploitation and training process. This oversight can lead to substantial overestimation of rewards, potentially leading to suboptimal solutions, particularly in the highly competitive environment of ride-pooling. Therefore, in this section, we present our novel Graph-based MARL algorithm, termed as GATDDQN, which effectively captures the agent interdependence with localized bipartite matching graph within a large-scale ride-pooling system. The rest of this section will proceed as follows. We will initiate our discussion by showing how to build upon the previous MDP framework to incorporate localized bipartite matching. We will then review the fundamentals of classical Graph Attention Neural Network techniques. Following this, we will delve into the structure and formulation of our GATDDQN, which is designed to capture agent interdependence through a localized bipartite matching graph.

\subsection{Localized Bipartite Match Interdependent MDP}
\begin{figure*}[ht]
  \centering
  \includegraphics[width=0.7\linewidth]{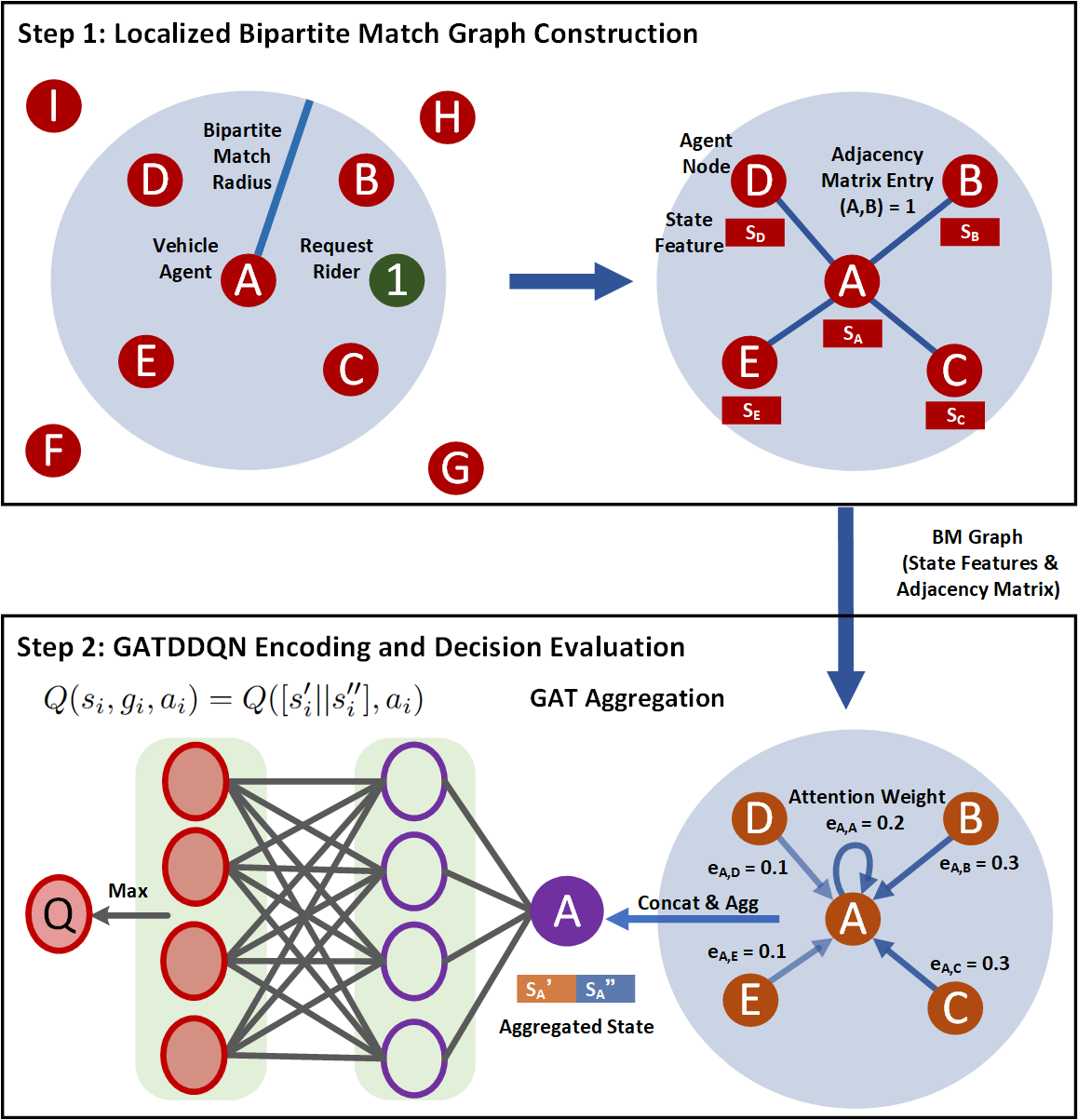}
  \caption{\blue{Visualization of GATDDQN algorithm pipeline. Step 1 constructs the localized bipartite match graph, passing the state features and adjacency matrix to Step 2. Step 2 performs GATDDQN and decision evaluation based on the inputs from Step 1.}}
  \vspace{-1.5em}
  \label{fig:GATS-MARL}
\end{figure*}

At time $t$, when coordinating vehicle agent fleets, the interdependence among vehicles primarily emerges from the order matching process. Agents within the pickup range of the same orders may encounter the same orders, leading to potential competition. With this in mind, for agent $n$, we can define a localized bipartite match graph $g_{n,t} = \{v_{n,t}, e_{n,t}\}$, where $v_{n,t}$ denotes the nodes representing agents within the localized graph, and $e_{n,t}$ denotes the edges. In such a graph, edges are drawn between the ego agent (refers to agent $n$ itself) and other agents only if those agents fall within a predefined proximity threshold. For the platform, as shown in Step 1 of Figure~\ref{fig:GATS-MARL}, we define an adjacency matrix where both the number of columns and rows correspond to the number of agents on the ride-pooling platform. A proximity threshold, referred to as the bipartite match radius, has been predefined. In this matrix, the entry $(i, j)$ is set to 0 if the distance between agent $i$ and agent $j$ exceeds this radius (like agent A and I in Figure~\ref{fig:GATS-MARL}), and to 1 if their distance is within the radius (like agent A and B in Figure~\ref{fig:GATS-MARL}). Consequently, the adjacency matrix for the localized bipartite match graph $g_{n,t}$ can be derived by referencing either the $n$-th row or column of the platform's matrix. Here, we define $\mathcal{N}(n)$ as the set of neighbors for agent $n$ within a certain radius, including the vehicle $n$ itself. By utilizing the localized bipartite match graph, we can refine the previously independent MDP model from Section III into a \textit{\textbf{localized bipartite match interdependent MDP}} model, where the $Q$ value is redefined accordingly:
\begin{equation}
\begin{aligned}
Q_{\text{tot}}(S_t, A_t) &= \sum_{n = 1}^{N} {\mathbb{E}_{\pi_{bm}} \left[ \sum_{k = 0}^{\infty} \gamma^k r_{n,t + k + 1} \mid s_{n,t}, g_{n,t}, a_{n,t} \right] }\\
&  = \sum\limits_{n = 1}^N {{Q}({s_{n,t}}, {g_{n,t}}, {a_{n,t}}} )
\end{aligned}
\end{equation}
where $\pi_{bm}$ represents the control policy for the localized bipartite match interdependent MDP. 
In this case, the goal of novel localized interdependent MDP model is to find the optimal policy $\pi^*_{bm}$ that maximizes the joint expected discounted cumulative reward over time:
\begin{equation}
Q^*_{\text{tot}}({S_t}, {A_t}) =  \sum\limits_{n = 1}^N {\mathbb{E}_{\pi^*_{bm}} \left[ {\sum\limits_{k = 0}^\infty  {{\gamma ^k}{r_{n,t + k + 1}}|{s_{n,t}},{g_{n,t}},{a_{n,t}}} } \right]} 
\end{equation}

\vspace{-1em}
\subsection{Classical Graph Attention Neural Network}
\blue{However, unlike the previous case of ride-sourcing~\cite{yang2018mean}, the challenge intensifies in ride-pooling dispatch, where fully understanding and aggregating the interdependence within the bipartite match graph becomes more complex. In a ride-pooling environment, each agent may have unique passengers with different itineraries. Traditional GCNs that employ average or max aggregation methods \cite{hamilton2017inductive} to learn interdependencies within a bipartite match graph can yield misleading or inaccurate information for decision-makers. Therefore, after constructing the localized bipartite match graph for each agent, we propose to employ GATs to enable an unassigned agent to dynamically weigh its neighbors based on the current scenario.}

To this end, we will first review the basic notations and ideas of classical GATs. In particular, GATs \cite{velivckovic2017graph} are designed to handle data structured as graphs $G = \{V,E\}$, where $V$ represents the nodes (which are agents in our context), and $E$ represents the edges. A single layer of GATs operates by computing a set of transformations and attention coefficients for each node in the graph. Firstly, for the message layer,  each node or agent $i$ in the graph is transformed using a shared linear transformation (e.g., GraphSAGE \cite{hamilton2017inductive}), parameterized by a weight matrix $W \in \mathbb{R}^{F' \times F}$:
\begin{equation} \label{eq:GAT_transform}
s_i' = W s_i
\end{equation}
where $s_i \in \mathbb{R}^F$ is the state of agent $i$ and $s_i' \in \mathbb{R}^{F'}$ is the transformed state vector.

For the aggregation layer, an attention mechanism computes attention coefficients $e_{ij}$ that capture the importance of agent $j$'s state to agent $i$ in its neighborhood $\mathcal{N}(i)$:
\begin{align} \label{eq:GAT_attention}
e_{ij} &= \frac{\exp(\sigma(a^T[s_i' || s_j']))}{\sum_{k \in \mathcal{N}(i)} \exp(\sigma(a^T[s_i'|| s_j']))} \\
& = \text{softmax}_j(\sigma(a^T [s_i' || s_j']))    
\end{align}
where $a: \mathbb{R}^{F'} \times \mathbb{R}^{F'} \rightarrow \mathbb{R}$ is a learnable shared attention mechanism, $[s_i' ||  s_j']$ denotes the concatenation of agent $i$ and $j$'s transformed states, and $\sigma$ is a non-linear activation function. Moreover, to stabilize the learning process and enrich model capacity, GATs employ multi-head attention as an aggregation function:
\begin{equation} \label{eq:GAT_aggregation}
s_i'' = \frac{1}{K} \left( \sum_{k \in \mathcal{K}} \sum_{j \in \mathcal{N}(i)} e_{ij}^k W^k s_j \right)
\end{equation}
where $K$ is the number of parallel attention mechanisms (heads), $e_{ij}^k$ is the normalized attention coefficient computed by the $k$-th attention head, $W^k$ is the corresponding weight matrix.

\subsection{GATDDQN for Localized Bipartite Match Graph}
In this subsection, we will combine GATs with the localized bipartite matching graph to capture the localized bipartite match interdependence of agents, leading to the GATDDQN in Step 2 of Figure~\ref{fig:GATS-MARL}. However, different from the simple attention mechanism in the preceding subsection, here for aggregation layer, we adopt transformer-style attention mechanism~\cite{vaswani2017attention} in Equation (\ref{eq:GAT_transformer_style_attention}) to compute the attention score in order to better capture and aggregate the complex information lying within the localized graph of ride-pooling system:
\begin{equation} \label{eq:GAT_transformer_style_attention}
e_{ij} = \text{softmax}_j \left( \frac{Q_i^\mathrm{T} K_j}{\sqrt{d_k}} \right)
\end{equation}
where \(Q_i = W^Q s_i'\), \(K_j = W^K s_j'\), and \(V_j = W^V s_j'\) with \(W^Q, W^K, W^V \in \mathbb{R}^{F' \times F'}\) being the weight matrices for queries, keys, and values, respectively, and \(d_k\) is the dimensionality of the key vectors for scaling. Similarly, we adopt multi-head attention 
to stabilize the learning process and enrich model capacity as follows:
\begin{equation} \label{eq:transformer aggregation}
s_i'' = W''(||_{k=1}^{K} \sum_{j \in \mathcal{N}(i)} e_{ij}^k W^k s_j)
\end{equation}
where $W''\in \mathbb{R}^{KF' \times F'}$ is the final linear transformation layer that maps the concatenation of the $K$ heads' aggregated features into the same dimensions of $s_i'$, and $||_{k=1}^{K}$ represents the operation of concatenating multiple heads' aggregated features (i.e., $\sum_{j \in \mathcal{N}(i)} e_{ij}^k W^k s_j$). 

With information aggregated, we could concatenate the state information of ego agent (i.e., $s_i'$) and aggregated states of the other agents in the localized Bipartite Match graph (i.e., $s_i''$) of last layer of GATs, and feed them into the downstream DDQN backbone structure introduced in section III, to achieve more collaborative multi-modal transportation behaviors among agents.  The whole flows are demonstrated in Figure \ref{fig:GATS-MARL}. The Q-value estimation towards state-action pair of agent $i$ at the time could then be represented by:
\begin{equation} \label{eq: GATDDQN action}
{Q}({s_{i}},{g_{i}},{a_{i}}) = Q([s_i'||s_i''],a_i)
\end{equation}


\section{BMG-Q: Effective, Scalable, and Robust Ride-Pooling Order Dispatch Framework}
With the ideas of localized bipartite match interdependent MDP  and GATDDQN backbone established, in this section we wil discuss how GATDDQN's value estimations could be combined with ILP via our proposed posterior score function for the bipartite match process to finally form the proposed BMG-Q framework. 

\subsection{Dynamic ILP via Posterior Score Function}
In existing literature, the ILP for the bipartite matching process typically relies on $Q(s,a,\theta)$, such as (\ref{eq:ILP})  in Section II. However, this approach encounters two issues: (1) it lacks an exploration mechanism in the bipartite matching process, which restricts the exploration of the vehicle agents' state space; (2) matching based solely on $Q(s,a,\theta)$ is prone to bias due to variability in the estimates, which can affect the decision-making process. \blue{These are particularly crucial in the case of ride-pooling vehicle dispatch, where ride-pooling requires agents now need to track detailed origin-destination (O-D) pairs and precise time detour information for each passenger in their states and thus have much larger state space compared with previous ride-sourcing setting. Therefore, sticking to conventional value function settings as in \cite{feng2022coordinating} will potentially lead to suboptimal solutions.} 

To deal with the above two issues and consider the importance of accurate value function~\cite{wei2022calibration}, here we propose posterior score function to better integrate the value function of GATDDQN with ILP. The formulation of our posterior score function $S(s_{n,t}, g_{n,t}, a_{n,t})$ is given in Equation (\ref{eq:Posterior Score Function}) below, with its visualization shown in Figure~\ref{fig:Dynamic ILP}. To effectively balance exploration and exploitation, we implement an $\epsilon$-greedy strategy and introduce a term $S_{\text{explore}}$ during the exploration phase. This term represents the upper bound of the Q-value, $S(s_{n,t}, g_{n,t}, a_{n,t})$, which is set to a significantly high value (e.g., 100,000) to encourage exploration in exploration stage:
\begin{equation} \label{eq:Posterior Score Function}
\begin{split}
S(s_{n,t}, g_{n,t}, a_{n,t}) = 
\begin{cases} 
  Q(s_{n,t}, g_{n,t}, a_{n,t})\\
 - b(s_{n,t}, g_{n,t}), & \text{with prob } 1 - \epsilon \\
S_{\text{explore}}, & \text{with prob } \epsilon
\end{cases}
\end{split}
\end{equation}
In the exploitation stage, to mitigate variance, we adjust the Q-value by subtracting a bias term $b(s_{n,t}, g_{n,t})$, which remains unaffected by the actions of the agents. This term can either be a constant or the state's value function, $V(s_{n,t}, g_{n,t})$. Employing $V(s_{n,t}, g_{n,t})$ gives rise to the advantage function $A(s_{n,t}, g_{n,t}, a_{n,t}) = Q(s_{n,t}, g_{n,t}, a_{n,t}) - V(s_{n,t}, g_{n,t})$. In this context, the advantage function $A(s_{n,t}, g_{n,t},  z)$ signifies the relative benefit of assigning agent $n$ to pick up a particular order $z$, hence quantifying the importance of the assignment.
Replaced with our score function, the novel dynamic ILP formulation can be written in Equation (\ref{eq:Dynamic ILP}) below:
\begin{equation} \label{eq:Dynamic ILP} 
\begin{aligned} 
& \underset{x_{n,z}}{\text{maximize}} 
& & \sum_{n=1}^{N} \sum_{z=0}^{Z_{t}} S(s_{n,t},g_{n,t},z) x_{n,z} \\ 
& \text{subject to} 
& & \sum_{n=1}^{N} x_{n,z} \leq 1, \quad \forall z, \\ 
& & & \sum_{z=1}^{Z_{t}} x_{n,z} \leq 1, \quad \forall n, \\ 
& & & x_{n,z} \in \{0,1\}, \quad \forall n,z \\
& & & \sum_{n=1}^{N} x_{n,z}\cdot d_{n,z}\leq R_{match}, \quad \forall z, \\
\end{aligned} 
\end{equation}
Compared with the ILP commonly adopted in the existing literature (\ref{eq:ILP}), the proposed score function not only captures the dependence of value estimations on the localized graph, but also captures the importance of order $z$ for vehicle $n$. We will show through numerical simulation that the proposed approach can significantly reduce overestimation and improve the overall performance of the proposed BMG-Q framework. 

\begin{figure}[ht]
  \centering
  \includegraphics[width=1\linewidth]{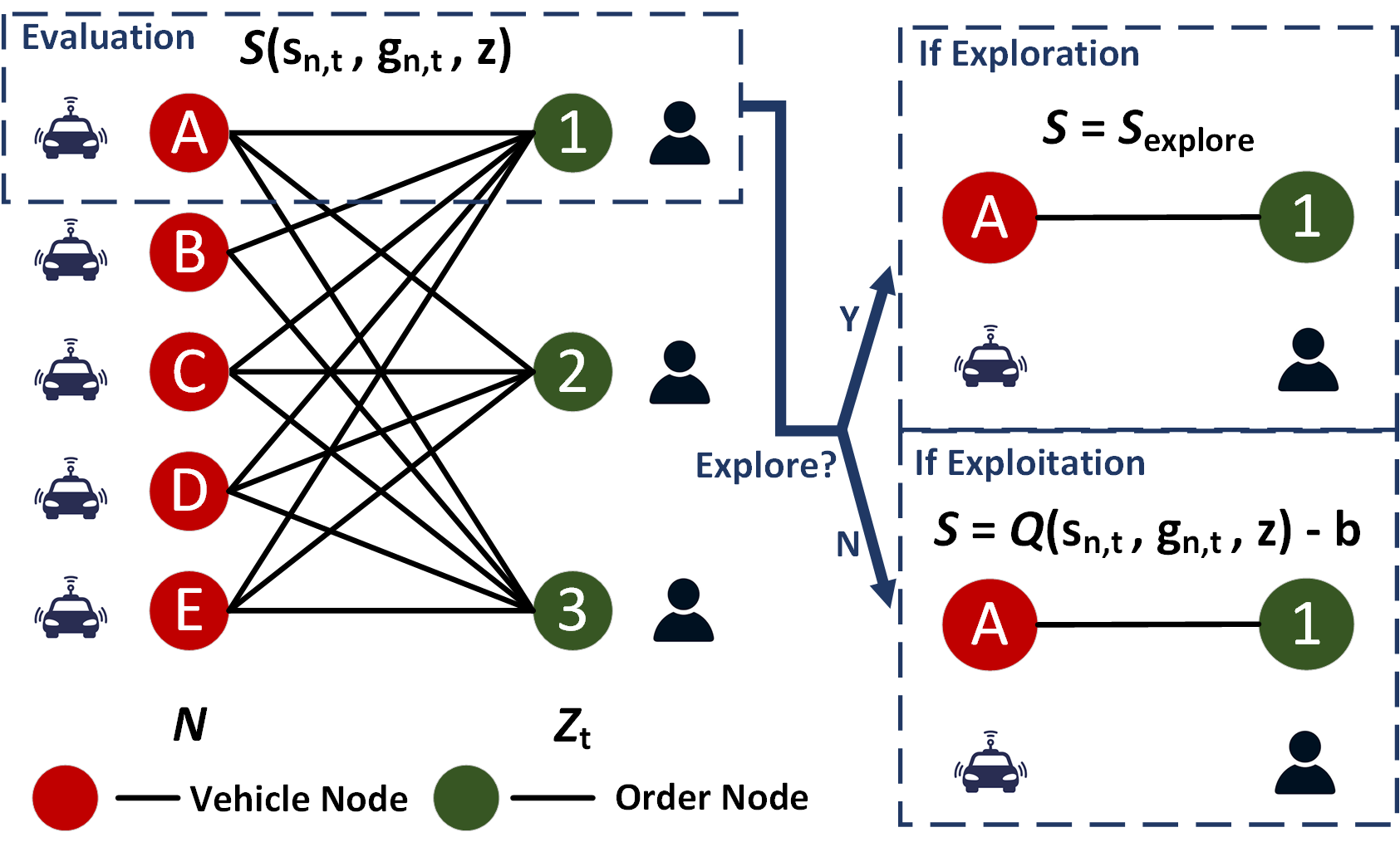}
  \caption{\blue{Dynamic ILP with posterior score function. Our proposed posterior score function $S(s_{n,t}, g_{n,t}, a_{n,t})$ combines (1) an exploration-exploitation trade-off strategy with (2) GATDDQN-derived advantage estimation to enhance matching decisions throughout both training and validation phases.}}
  \vspace{-1em}
  \label{fig:Dynamic ILP}
\end{figure}


\subsection{Training GATDDQN for Large-Scale System}
Since GATs could be trained with downstream neural network loss functions, the GATDDQN backbone could be trained end to end via slightly modifying the TD error introduced in Equation (\ref{eq:DDQN Loss Calculation}) into the Equation (\ref{eq:GATDDQN Loss Calculation}) below.
\begin{equation} \label{eq:GATDDQN Loss Calculation}
\begin{aligned}
L = \mathbb{E}_{\tau \sim \mathcal{D}}\Bigg[&\Bigg{(}r_i + \gamma Q\left(s_i', g_i',\mathop{\arg\max}_{a_i'} Q(s_i',g_i',a_i';\theta);\theta^- \right) \\
&- Q(s_i,g_i,a_i;\theta)\Bigg{)}^2\Bigg] 
\end{aligned}
\end{equation}
where $Q(s,g,a;\theta)$ is the Q value estimated by the training network whose neural network parameter is $\theta$ and $Q(s,g,a;\theta^-)$ is the Q value estimated by the target network $\theta^-$, $\tau$ is the trajectory from the sampled mini-batch $\mathcal{D}$. Subsequently, we could update parameters of training network and target network with Equations (\ref{eq:DDQN update}) and (\ref{eq:Polyak Average}) respectively. 

However, during our implementation of GATDDQN into very large-scale system with thousands of agents, we find that training could potentially become unstable due to shift of dynamic graphs and agents' over-fitting to recent experience tuples \cite{jiang2018graph,liu2020multi,kullman2022dynamic,zhang2022dynamic}. To address with this issue, we propose and adopt two simple but effective techniques:

Firstly regarding the shift of dynamic graph in large systems, as all agents learn simultaneously within the environment, localized bipartite graph representations (such as number and states of neighboring agents) could change dramatically from one time window to another in the training phase. This spatial-temporal shift poses significant challenges for GNN encoding and learning~\cite{jiang2018graph,liu2020multi,zhang2022dynamic}. To mitigate this gap, we propose graph sampling strategy, which is implemented prior to inputting each agent's bipartite match graph into the GAT. The strategy first involves sampling a fixed number of agents when the number of agents in the bipartite match graph exceeds this predetermined threshold. For instance, if the fixed number is set to 30 neighboring cars, and a vehicle agent has 50 vehicle agents nearby, then the agent will only randomly consider 30 of them. Conversely, if the number of agents in the bipartite match graph is fewer than the fixed amount, we introduce dummy nodes to maintain a constant graph size. For example, if a vehicle has only 10 neighboring vehicle agents, we will add 20 dummy nodes (represented as zero vectors) to the bipartite graph. With this graph sampling strategy, the GAT aggregator at each decision epoch consistently considers and encodes a fixed bipartite graph topology of 30 nodes during training. This approach not only helps to stabilize the training and improve training efficiency by reducing the state space variability but also preserves the generality of the model, ensuring that the training remains effective across different scenarios.

Secondly, to deal with the problem of over-fitting to recent experience tuples, we adopt gradient clipping to Equation (\ref{eq:DDQN update}) as in Equation (\ref{eq:Gradient Clipping}), where $||\cdot||_2$ stands for L-2 norm:
\begin{equation}\label{eq:Gradient Clipping}
\begin{aligned}
g &= \Delta_\theta L, \\
g_{\text{clip}} &= \begin{cases}
g \times \frac{\text{threshold}}{||g||_2}, & \text{if } ||g||_2 > \text{threshold}, \\
g, & \text{otherwise},
\end{cases} \\
\theta &= \theta - \alpha g_{\text{clip}}.
\end{aligned}
\end{equation}
The gradient clipping enforces the agents to update the policy within a region so as to not over-fit to recent experience tuples and guarantee to improve its policy. Through our further training and validation, we find that the two tricks not only help to stabilize the training process but also make the framework more robust to task variability and parameter change.

\subsection{Summary of the proposed BMG-Q Framework}
Finally, we give a summary of our whole BMG-Q Framework. The framework could be visualized using Figure \ref{fig:Overview of BMG-Q} below. Specifically, during each time window, when new orders arrive, unmatched orders and vehicle information are first re-sorted and updated. With the evaluations from the GATDDQN network, the central platform assigns these orders to vehicle agents via solving the ILP. After bipartite match assignments, the vehicle agents perform their respective actions and collect their experiences for further GATDDQN learning. Subsequently, the routing system updates the routes and estimated times of arrival (ETAs), which are then communicated back to the central platform.

The training details of our BMG-Q framework could be found at Algorithm \ref{al: BMG-Q Framework}. Specifically, after initializing the simulator and GATDDQN in steps 1 through 4, we enter the training phase. To achieve a balance between exploitation and exploration, we perform exploration decay to exploration rate of the bipartite matching process in step 5. This gradual reduction in exploration rate is designed to  transition the focus from exploration to exploitation as the learning advances. In steps 9 to 12, similar to the DDQN backbone of ILPDDQN, our GATDDDQN backbone adopts double networks, experience replay, and soft update. Thanks to the localized bipartite match graph topology, graph sampling, and gradient clipping introduced in steps 9 to 11, GATDDQN backbone manages to learn to capture the localized interdependence in very large-scale system with thousands of agents, thus leading to more effective assignment decisions of the overall BMG-Q framework.

\begin{algorithm}
\caption{BMG-Q Framework} \label{al: BMG-Q Framework}
\begin{algorithmic}[1] 
\STATE Simulator Initialization: OSRM Router Model\cite{ProjectOSRM}, Number of Vehicles $N$, Matching Distance $R_{match}$ 
\STATE GATDDQN Initialization: Memory $M$, Memory Capacity $C$, Training Net Parameter $\theta$, Target Net Parameter $\theta^-$, and Training Hyper-parameters $\alpha$, $\rho$, $\text{threshold}$, Exploration Rate $\epsilon$, $\epsilon_T$ with Exponential Decay Rate $\beta$.
\FOR{$e = 1$ to Episodes}
    \STATE Initialize: Episode Order Requirements, and Number of Vehicles $N$
    \STATE Perform exponential decay according to Equation (\ref{eq:epsilon_decay})
    \FOR{$t = 0$ to $t_{\text{terminal}}$ by $\Delta t$}
        \STATE Central platform updates order information, each vehicle’s location, and on-board passenger situations.
        \STATE Central platform assigns orders to vehicle agents according to Score Function and ILP formulation in Equations (\ref{eq:Posterior Score Function}) and (\ref{eq:Dynamic ILP}).
        \STATE Vehicles observe their orders, perform their assigned actions in the simulation platform and add every agent’s new experience tuple $(s, g, a, r, s', g')$ into the memory $M$.
        \IF{memory size larger than $C$}
            \STATE Sample $N$ experience tuples $(s, g, a, r, s', g')$ in $M$ as mini-batch $D$ and use Equation~(\ref{eq:GATDDQN Loss Calculation}) and~(\ref{eq:Gradient Clipping}) to update $\theta$.
            \STATE Update target network parameters $\theta^-$ using Equation~(\ref{eq:Polyak Average}).
        \ENDIF
        \STATE Based on the chosen action, central platform calculates the new route and estimated time of pickup and drop off.
    \ENDFOR
\ENDFOR
\end{algorithmic}
\end{algorithm}

\begin{figure*}[ht]
  \centering
  \includegraphics[width= 1 \linewidth]{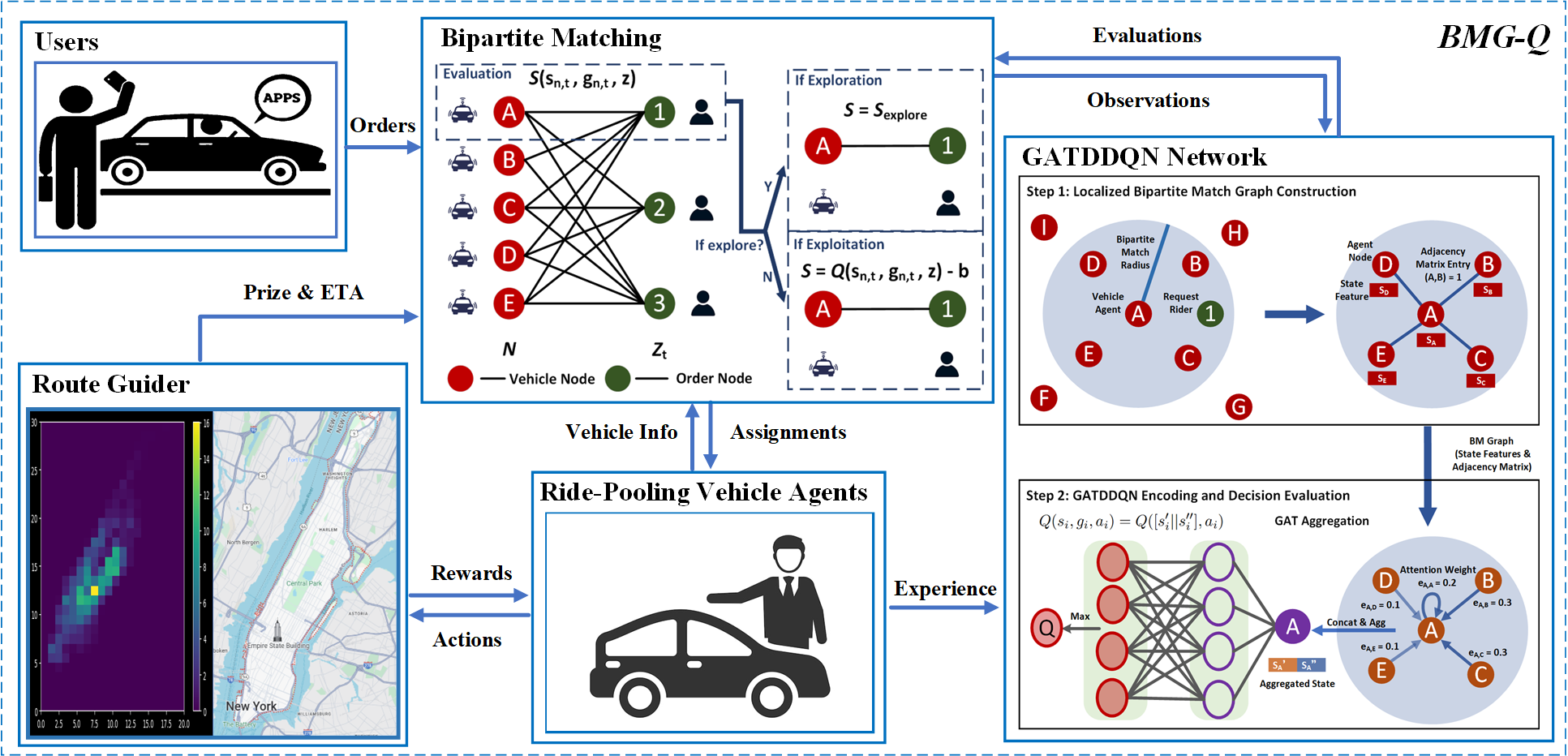}
  \caption{\blue{Overview of the proposed BMG-Q framework for ride-pooling vehicles dispatch. For every decision round, initially users submit orders through a mobile application, and the system updates and sorts these orders alongside vehicle information. Taking into account long-term uncertainties including the intricate interdependence of agents, the GATDDQN network evaluates and dynamically assigns orders to suitable vehicles using ILP. After assignments, vehicle agents execute their actions, with experiences collected for subsequent learning phases of the GATDDQN network. Concurrently, the Route Guider updates routes and estimated times of arrival, which are communicated back to the central platform. (Part of icons and map are from \cite{GoogleImages, GoogleMaps}).}}
  \vspace{-1em}
  \label{fig:Overview of BMG-Q}
\end{figure*}

\section{Case Studies}
This sections presents a  case study that utilizes real-world data from Manhattan, New York City. We will first detail the implementation of our simulation framework, after which we will showcase the effectiveness, scalability and robustness of our BMG-Q framework through the training and validation results.

\subsection{Simulation Setup}
\begin{figure}
  \centering
  \includegraphics[width=0.75\linewidth]{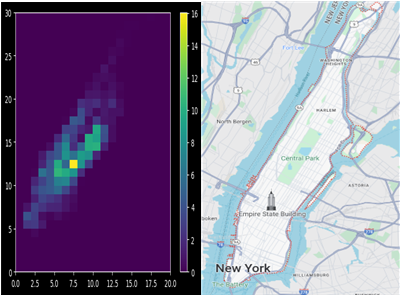}
  \caption{Demand zone visualization (Map is from \cite{GoogleMaps})}
  \vspace{-1.5em}
  \label{fig:Demand Visualization}
\end{figure}

The simulation environment for this study is based on the public dataset of taxi trips in Manhattan, New York City~\cite{al2019deeppool,NYCTaxiData2018}. This dataset includes detailed information for each trip, such as pickup and dropoff times, origin and destination geo-coordinates, trip distance, and duration. Focusing on peak hours—specifically from 8:00 AM to 10:00 AM—we tailored the training dataset to include data from trips that occurred between 8:00 AM and 8:30 AM on May 4, 2016. During this half-hour period, the average order density reached approximately 275 trips per minute in central Manhattan, totaling about 8,250 orders. For the validation dataset, we similarly extracted data from trips within the same half-hour window but on various days throughout May 2016. We divided Manhattan into 57 zones. This zoning was informed by the distribution of orders and a resolution of 800m x 800m was used, for which a visualization is shown in Figure~\ref{fig:Demand Visualization}. To serve these demand with a minimum service rate of 85\% across all RL methodologies tested in this study, we set the number of ride-pooling agents as 1000 and number of vacant seats of each vehicle as 3. To provide real-time route guidance and estimating the passengers' onboard time, we employed the OSRM model~\cite{ProjectOSRM} through docker as our router. The coefficients of reward function is set as ${\beta _0} = 100$, ${\beta _1} = 40$, ${\beta _2} = 5$, ${\beta _3} = 2$, ${\beta _5} = 20$, ${thre} = 15$, with the aim to encourage agents to pick up more orders but not result in large average detours of passengers. For bipartite match process, we set matching distance $R_{match}$ as 1.2 km and any requests that remain unmatched with vehicles for more than five minutes will be automatically rejected. Additionally, it's important to note that our focus is on fully optimizing the potential of the ride-pooling fleet order dispatch during peak hours, when the demand is high and the occupancy of the vehicles are naturally high. Therefore, we have chosen not to include rebalancing operations \cite{alonso2017demand} in our approaches or any of the benchmark approaches considered in this study. The incorporation of rebalancing operations is left for future work.

In implementing GATDDQN, we adopted a linear transformation as message passing layer and multi-head attentions with a head count of 3 as aggregation layer to form the backbone structure of GATs.  For the sake of a receptive field that is manageable \cite{xu2018powerful} and to maintain simplicity, we have opted to employ a single-layer GAT. Complementing this, we used a Multi-layer Perceptron (MLP) with a three-layer configuration and RELU as the non-linear activation function to establish the neural network backbone of DDQN. \blue{The hidden unit of MLP is set as 256 and MLP's output signifies the Q-value for the distinct actions of GATDDQN.} For further extension to scenarios like passenger transfer\cite{feng2022coordinating} and relocations\cite{al2019deeppool}, these actions could be further expanded to cover the decision of whether to refrain from picking up a potential new passenger or to indeed embark the passenger within one of the zones or stations on the 57-zone map. With the vehicle capacity set to three, the input state into the GATs constitutes a 1-by-14 tensor representing each vehicle's state, while the final aggregated input state to DDQN is a 1-by-128 tensor. We set the memory capacity $C$ at 20,000 and the mini-batch size of $D$ as 1024. For training updates, further technical considerations included employing the Mean Squared Error (MSE) loss and utilizing Adam as the optimizer for GATDDQN, with an assigned learning rate $\alpha = 0.01$, soft update rate $\rho = 0.005$, and gradient clipping threshold as 0.05. For graph sampling, we set the fixed number of agents as 30 and observation distance as equal to bipartite match distance $R_{match}$. Regarding exploration and exploitation trade-offs, we set the initial exploration rate $\epsilon$ as 1, exploration decay rate $\beta =0.996$, and exploration final value $\epsilon_T$ as 0.005. For every training in the following sessions, we standardized the comparison by configuring representative frameworks with neural network architectures and hyper-parameters that closely mirror those used in our BMG-Q model, and train our BMG-Q and representative frameworks for around 2000 episodes on Intel 14700K CPU and NVIDIA GEFORCE 4080 GPU desktop setup\footnote{Although the training process may take up to two days to complete, our BMG-Q is efficient for real-time decision-making at scale. Once the BMG-Q is trained, for each minute of dispatch decisions involving approximately 1000 vehicles and 300 orders, the dispatch process takes only about 1 to 2 seconds on our desktop.}. 



\subsection{Effectiveness and Scalability of BMG-Q Framework}
Firstly, to test the effectiveness of BMG-Q framework in training, we compare our framework with three representative frameworks from modelling and MARL in ride-pooling: Greedy \cite{alonso2017demand,al2019deeppool},  ILP + Independent RL \cite{al2019deeppool,sadeghi2022reinforcement, feng2022coordinating, wang2023optimization}, ILP + Independent RL Considering Agents Nearby \cite{shah2020neural}, and vanilla ILP + Attention-based MARL \cite{kullman2022dynamic,enders2023hybrid}. For Greedy framework, we train a reward model offline till convergence and replace the Q function $Q(s,a)$ in Equation (\ref{eq:ILP}) with reward function $r(s,a)$. For ILP + Independent framework, we adopt ILPDDQN given in Algorithm \ref{al: ILPDDQN Framework} and also add exploration inside for the aim of fair comparisons. For ILP + Independent RL Considering Agents Nearby, we extend ILPDDQN baseline by incorporating
the count of other agents and requests within its current zone into the agent's state representations, termed as IQL\_CAN. For vanilla ILP + Attention-based MARL, we remove our localized graph, gradient clipping and graph sampling strategy, which however lead the training to become significantly unstable in large-scale system with 1000 agents (thus not shown in the simulation results). 

\begin{figure}
  \centering
  \includegraphics[width=0.9\linewidth]{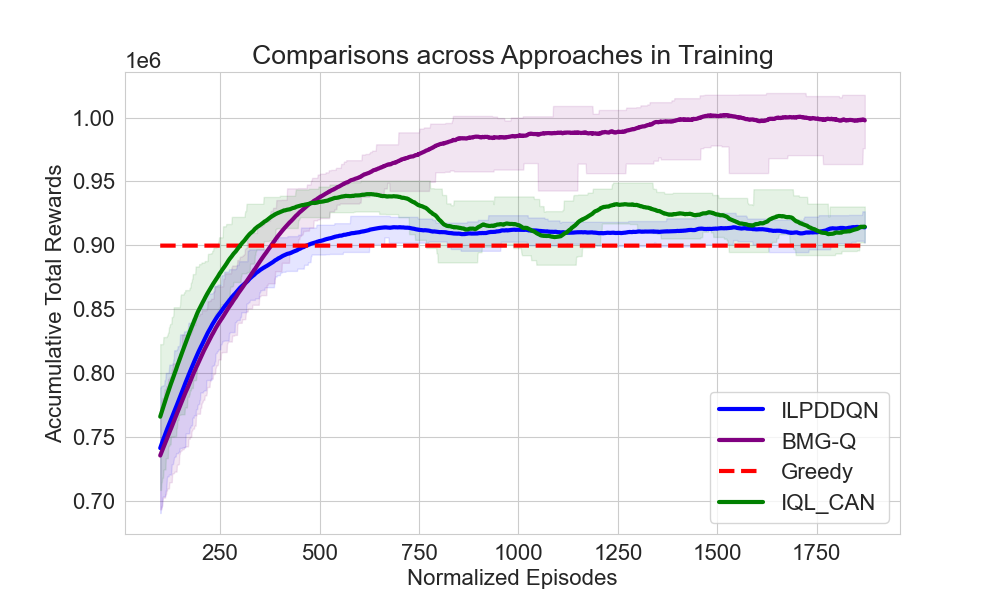}
  \caption{Training comparison across different approaches}
  \vspace{-1em}
  \label{fig:Training Comparison across Different Approaches}
\end{figure}

The training curves are shown in Figure \ref{fig:Training Comparison across Different Approaches}. As we could observe, our BMG-Q framework performs significant better than other three baselines, with respect to accumulative total rewards and training stability. Firstly, the BMG-Q curve (purple) demonstrates a rapid ascent early in the training process and achieves higher accumulative reward values than all the other approaches. Moreover, the stability of the BMG-Q approach is evident from the relatively tight confidence interval (shaded purple area) which indicates less variation in the performance across different training runs. This contrasts particularly with the IQL\_CAN approach (green), which, despite improving over time, shows a broader confidence interval, implying more variability in its performance.

To demonstrate the superiority of our BMG-Q framework in terms of transportation benefits, we selected several key transportation metrics for evaluation, including service rate, average passenger waiting time, average travel detour, and vehicle kilometers traveled. We evaluated the results using trip data from Wednesday. Our comparison of these selected metrics under the proposed BMG-Q learning framework against three benchmark algorithms is presented in Table~\ref{tab:metric_comparison}. The results clearly indicate that BMG-Q outperforms the baseline methods across the following metrics: cumulative total rewards, average passenger waiting time, service rate, and vehicle kilometers traveled. Specifically, when compared to IQL\_CAN, ILPDDQN, and the Greedy algorithm, the accumulated total reward improved by 8.9\%, 11.8\%, and 11.9\%, respectively; the service rate increased to 94.4\%, up from 85.4\%, 85.6\%, and 86.5\%, respectively; the average passenger waiting time was reduced by 37.5\%, 38.2\%, and 1.8\%, respectively; and the vehicle kilometers traveled were reduced by 0\%, 2.2\%, and 7.7\%, respectively. However, we note that the average travel detour of BMG-Q is 25.8\% larger than that of the ILPDDQN. This can be intuitively explained since the proposed algorithm enables a higher chance of matching and higher service rate, which naturally leads to slightly more average travel detours as a consequence.

\begin{table*}[ht!]
  \centering
  \caption{Comparison across Approaches in Terms of Transportation Metrics}
  \begin{tabular}{lcccc}
    \toprule
    Metric & BMG-Q & IQL\_CAN & ILPDDQN & Greedy \\
    \midrule
    Accumulative Total Reward ($\times10^6$) & \textbf{1.007} & 0.925 & 0.901 & 0.900\\
    Service Rate & \textbf{94.4\%} & 85.4\% & 85.6\% & 86.5\% \\
    Average Passenger Waiting Time & \textbf{2.17 min} & 3.47 min & 3.51 min & 2.21 min \\
    Average Travel Detour & 3.17 min & 2.69 min & \textbf{2.52 min} & 3.69 min \\ 
    Vehicle Kilometers Traveled & \textbf{13.1 km} & 13.1 km & 13.4 km & 14.2 km \\
    \bottomrule
  \end{tabular}
  \label{tab:metric_comparison}
\end{table*}

\begin{figure}[h]
  \centering
  \includegraphics[width=0.9\linewidth]{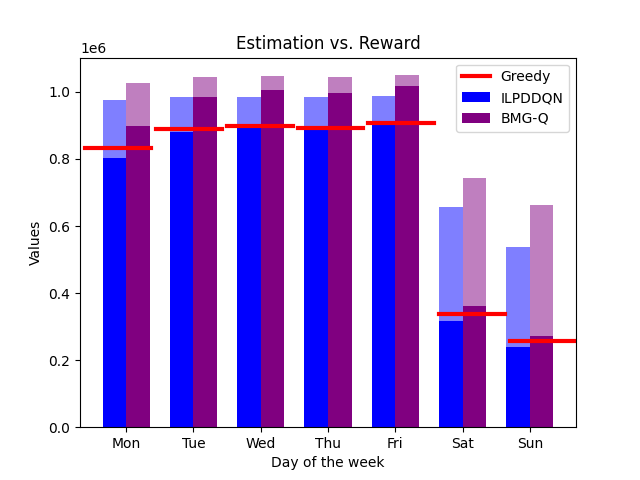}
  \caption{Validation comparison of BMG-Q across one week}
  \label{fig:Validation Comparison of BMG-Q across 1 Week}
\end{figure}

To further understand the rational why our BMG-Q manages to significantly outperform ILPDDQN, we validate BMG-Q, ILPDDQN, and Greedy using the data across an entire week. The comparison between agent's estimation and total accumulative rewards for 1000 cars is given in Figure \ref{fig:Validation Comparison of BMG-Q across 1 Week}. For each bar in Figure \ref{fig:Validation Comparison of BMG-Q across 1 Week}, the darker shade represents the actual reward, while the lighter shade indicates the amount of overestimation. As observed, the ILPDDQN's performance is hindered by significant overestimation, stemming from a complete disregard for potential interdependencies.  Consequently, while ILPDDQN still manages to slightly outperform the Greedy approach when validated on days similar to the one trained on (e.g., Thursday and Friday following a Wednesday training), its effectiveness diminishes on markedly different days like Monday, Tuesday, Saturday, and Sunday. On these days, the overestimation issue prevents ILPDDQN from accurately capturing task variations, leading to poorer performance compared to the Greedy Baseline. In contrast, our BMG-Q framework successfully mitigates the overestimation by more than 50\%, which leads to an impressive performance improvement compared to ILPDDQN.

\begin{table*}[t!]
  \centering
  \caption{Validation of BMG-Q and benchmark across various fleet agent number shifts}
  \label{tab: Validation of BMG-Q and Benchmark Across Various Fleet Agent Number Shifts}
  \begin{tabular}{lcccccc}
    \toprule
    & \multicolumn{3}{c}{BMG-Q} & \multicolumn{3}{c}{ILPDDQN} \\
    \cmidrule(lr){2-4} \cmidrule(lr){5-7}
    Metrics & 800 Cars & 1000 Cars & 1200 Cars & 800 Cars & 1000 Cars & 1200 Cars \\
    \midrule
    Rewards & 869,298 & 1,006,925 & 1,035,778 & 814,968 & 900,793 & 915,003 \\
    Order Pickup & 6,698 & 7,785 & 7,994 & 6,358 & 7,063 & 7,107 \\
    Passenger Detour (mins) & 3.32 & 3.17 & 3.01 & 2.75 & 2.52 & 2.36 \\
    \bottomrule
  \end{tabular}
\end{table*}

\begin{figure}[h]
  \centering
  \includegraphics[width=0.9\linewidth]{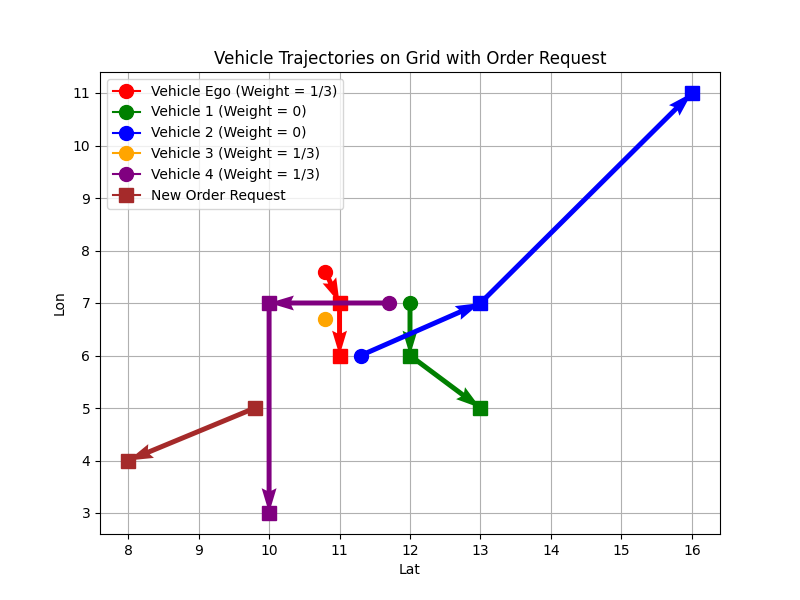}
  \caption{Illustrate example of localized graph attention excerpted from simulation results. The brown square represents a new order request. The circles in various colors correspond to different vehicles, and the squares in corresponding colors indicate the passengers already on board these vehicles. The `ego vehicle' (red dot), carrying two passengers, evaluates this request against neighboring agents. Priority is given to agents 3 (orange) and 4 (purple) due to potential route compatibilities, each with a weight of $1/3$. Agent 1 (green) and Agent 2 (blue) are ignored due to conflicting passenger trajectories.}
  \label{fig:Demonstration of Localized Graph Attention}
  \vspace{-1.5em}
\end{figure}

Additionally, using the training parameters specified for GATDDQN, we have drawn an illustrative example obtained from the simulation, as shown in Figure \ref{fig:Demonstration of Localized Graph Attention}, to examine how our BMG-Q framework discerns the intricate interdependencies within the bipartite matching graph. In this example, the brown square represents a new order request. The circles in various colors correspond to different vehicles, and the squares in matching colors indicate the passengers already on board these vehicles. The `ego vehicle', marked by a red dot and carrying two passengers, is assessing an order request, denoted by a brown square, alongside neighboring agents labeled 1 through 4. During the graph attention aggregation phase, the `ego vehicle' prioritizes agents 3 in orange and 4 in purple (with weights of 1/3 each). This prioritization is because, for agents 3 and 4, accepting the brown order does not conflict with the routes of their onboard passengers. Conversely, agents 1 in green and 2 in blue are disregarded by the `ego vehicle' (assigned weights of 0 respectively) because the brown order would interfere with the trajectories of their current passengers. Consequently, potential competition for the brown order arises primarily between the `ego vehicle' and agents 3 and 4. The training and validations results above are consistent with the intuition and prove the effectiveness of our proposed BMG-Q framework in large-scale ride-pooling order dispatch.

\subsection{Robustness of BMG-Q Framework to Task Variations}
\begin{figure}[h]
  \centering
  \includegraphics[width=0.9\linewidth]{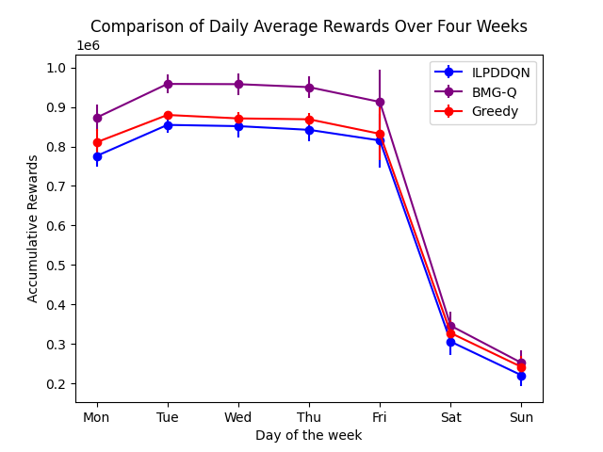}
  \caption{Validation of BMG-Q across task variations}
  \label{fig:Validation of BMG-Q across Task Variations}
  \vspace{-2em}
\end{figure}

In practice, the trained neural network may have to be applied to distinct scenarios, such as varying market conditions and fleet sizes. To validate the robustness of our BMG-Q framework under distinct scenarios, we first train the neural network on a specific scenario—peak hours on a Wednesday using a fleet of 1000 cars, and then test it across a range of fleet sizes and task variations. Specifically, we first explore its adaptability to different fleet size configurations of 800, 1000, and 1200 vehicles during the same time period, as presented in Table \ref{tab: Validation of BMG-Q and Benchmark Across Various Fleet Agent Number Shifts}. The results demonstrate that BMG-Q outperforms the ILPDDQN model consistently across multiple metrics, including rewards and order pickups, regardless of the fleet size. Subsequently, we extended our evaluation of the BMG-Q framework to check its performance across task variations over an entire month, as depicted in Figure \ref{fig:Validation of BMG-Q across Task Variations}. This comparison sheds light on the framework's robustness against fluctuating operational conditions with 1000 vehicles. It was observed that the BMG-Q framework consistently outstripped both the ILPDDQN and Greedy baselines in terms of daily average rewards over the span of four weeks. From these two sets of validation exercises, we can infer that our BMG-Q framework demonstrates robustness compared to previous benchmarks not only to variations in the fleet size but also to task variations common in ride-pooling scenarios of TNCs, such as varying fleet sizes, and day-to-day policy adaptation.  


Furthermore, as pointed out in \cite{alonso2017demand,shah2020neural}, ride-pooling vehicle fleets could have different number of seats settings in real-world. Accordingly, we retrained our BMG-Q framework for a fixed fleet size of 1,000 ride-pooling vehicles, adjusting vehicle capacities to 5, 8, and 10 seats. We also accounted for differences in operational costs, set at 0.1 per seat per minute. The simulation results, compared to the 3-seat setting, are presented in Table \ref{tab:seats_comparison}. The results reveal some trade-offs in selecting vehicle capacity. As seat capacity increases, the service rate improves, and average passenger waiting time is reduced. However, this comes at the cost of increased average travel detours and higher operational costs. As a consequence of this trade-off, the maximum reward is obtained when the seat capacity is equal to 5 (although the result is very close to that with a capacity 3). The result also show that our BMG-Q could be effectively trained on tasks with varying numbers of seats, finding efficient policies that maximize the potential of ride-pooling fleets.

\begin{table*}[t!]
  \centering
  \caption{BMG-Q Training Results for 1000 Ride-pooling Vehicles with Different Number of Seats}
  \begin{tabular}{lcccc}
    \toprule
    Metrics & 3 Seats & 5 Seats  & 8 Seats & 10 Seats\\
    \midrule
    Accumulative Total Reward ($\times10^6$) & 1.000 & 1.010 & 0.980 & 0.974\\
    Service Rate & 94.4\% & 96.9\% & 97.6\% & 98.8\% \\
    Average Passenger Waiting Time & 2.17 min & 1.98 min & 1.92 min & 1.73 min \\
    Average Travel Detour & 3.17 min & 3.15 min & 3.25 min & 3.36 min \\ 
    Vehicle Kilometers Traveled & 13.1 km & 12.9 km & 13.1 km & 13.1 km \\
    \bottomrule
  \end{tabular}
  \label{tab:seats_comparison}
\end{table*}

\subsection{Sensitivity Analysis of BMG-Q}

\begin{figure}[h]
  \centering
  \includegraphics[width=0.9\linewidth]{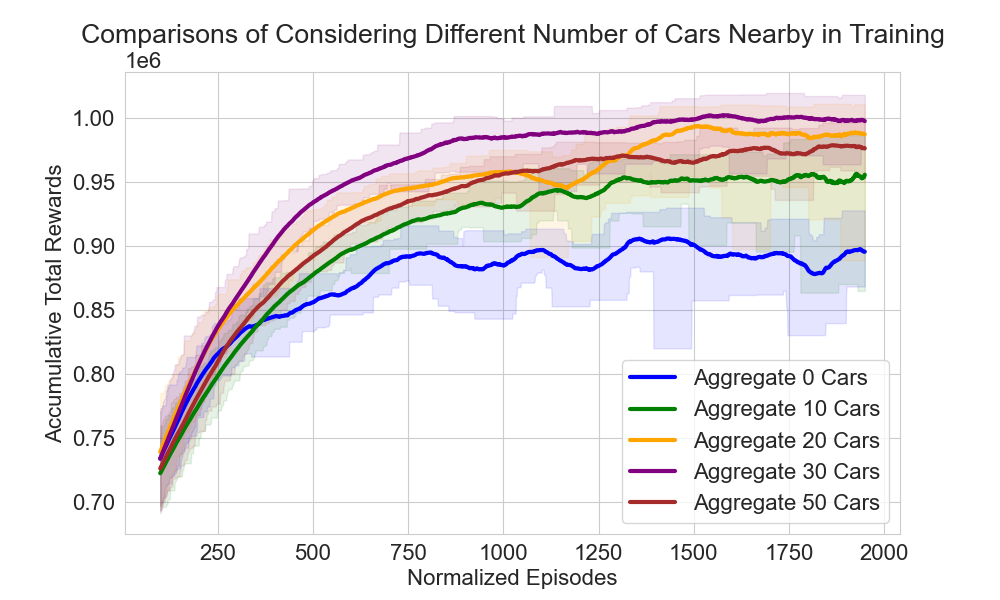}
  \caption{Graph sampling of different cars nearby in BMG-Q training}
  \label{fig:Graph Sampling of Different Cars Nearby in BMG-Q Training}
  \vspace{-1em}
\end{figure}

To evaluate the sensitivity of our proposed BMG-Q framework with respect to training hyperparameters, we conducted a comparative analysis of the training performance when varying the number of neighboring vehicles sampled for each ego vehicle. Specifically, we trained the neural network on ride-hailing data from a Wednesday scenario involving 1000 cars, with the results illustrated in Figure \ref{fig:Graph Sampling of Different Cars Nearby in BMG-Q Training}. It is noteworthy that our training framework remains stable during the training phase in a large-scale system, thanks to our gradient clipping and graph sampling strategy, and this stability is maintained irrespective of the number of neighboring vehicles sampled. Furthermore, we interestingly discovered that once the number of sampled vehicles reaches a certain threshold, further increasing the sample size does not significantly impact the training performance (e.g., the brown curve vs. the purple curve), which validates that the use of localized bipartite graph while sampling a limited number of neighboring vehicles can well capture the interdependence between agents. 

\section{Conclusion}
This paper proposes the Localized Bipartite Match Graph Attention Q-Learning (BMG-Q), a novel effective, scalable, and robust MARL algorithm framework tailored for large-scale ride-pooling order dispatch. By integrating localized bipartite match within the MDP of the ride-pooling system, we developed GATDDQN as a novel MARL backbone to accurately capture the dynamic interactions among agents in the large-scale ride-pooling order dispatch systems. Enhanced by gradient clipping and localized graph sampling, our GATDDQN improves scalability and robustness for very large-scale system, while the inclusion of a posterior score function in ILP captures the online exploration-exploitation trade-off and assists to reduce potential overestimation bias of agents. Through extensive experiments and validation, we show that BMG-Q demonstrates a superior performance in both training and operations of thousands of vehicle agents, outperforming benchmark RL frameworks by around 10\% in accumulative rewards and showing a significant reduction in overestimation bias by over 50\% while maintaining robustness and effectiveness amidst task variations and fleet size changes. \blue{Future extensions of our BMG-Q framework could explore applications in multimodal transportation systems~\cite{hu2025coordinating}, power system operations~\cite{qu2022scalable}, and other spatial-temporal decision games~\cite{WANG2025112711}, leveraging its inherent strengths in handling diverse network topologies and multi-agent interdependence across these domains.}

\bibliographystyle{IEEEtran}
\bibliography{base}

\vfill

\end{document}